\documentclass{amsart}  
\usepackage{amssymb}  
\usepackage{latexsym}  
  
\newcommand{\Z}{{\mathbb Z}}  
\newcommand{\R}{{\mathbb R}}  
\newcommand{\C}{{\mathbb C}}  
\newcommand{\N}{{\mathbb N}}

\newtheorem{theorem}{Theorem}     
     
\newtheorem{lemma}{Lemma}[section]     
\newtheorem{prop}[lemma]{Proposition}     
     
\newtheorem{definition}[lemma]{Definition}     
\newcommand{\tr}{{\mathrm{tr}}}  
  
\newcounter{smalllist}

\begin{document}  
\title[Dynamical Upper Bounds for One-Dimensional Quasicrystals]{Dynamical Upper Bounds for One-Dimensional Quasicrystals}  
\author{David Damanik}  
\thanks{Research partially supported by NSF Grant DMS--0010101} 
\maketitle  
\vspace{0.3cm}      
\noindent      
Department of Mathematics 253--37, California Institute of Technology,      
Pasadena, CA 91125, U.S.A.\\[2mm]      
E-mail: \mbox{damanik@its.caltech.edu}\\[3mm]      
2000 AMS Subject Classification: 81Q10, 47B80, 68R15\\      
Key words: Schr\"odinger operators, quasiperiodic potentials, quantum dynamics  
\begin{abstract}  
Following the Killip-Kiselev-Last method, we prove quantum dynamical upper bounds for discrete one-dimensional Schr\"odinger operators with Sturmian potentials. These bounds hold for sufficiently large coupling, almost every rotation number, and every phase.
\end{abstract}  
  
\section{Introduction}
In this paper we study quantum dynamical properties of one-dimensional quasicrystals. That is, we will be concerned with the family of operators $H_{\lambda,\alpha,\theta}$ on $\ell^2(\Z)$, acting on $u \in \ell^2(\Z)$ by  
    
\begin{equation}\label{oper}
H_{\lambda,\alpha, \theta} u(n) = u(n+1) + u(n-1) + \lambda v_{\alpha, \theta}(n) u(n).   
\end{equation}   
Here, $\lambda > 0$, $\alpha \in (0,1)$ is irrational, $\theta$ belongs to $ [0,1)$, and $v_{\alpha,\theta}(n)$ is given by  

\begin{equation}\label{potform}
v_{\alpha,\theta}(n) = \chi_{[1-\alpha,1)}(n \alpha + \theta \mod 1).
\end{equation}

Operators of the form \eqref{oper} have been extensively studied since the early eighties. We refer the reader to \cite{d2,s2} for history and known results (``last-millennium results''). The spectral type has been found to always be purely singular continuous \cite{bist,dkl}. Moreover, in some cases the spectral measures were even found to be absolutely continuous with respect to certain Hausdorff measures \cite{d1,dkl,dl,jl2,kkl} from which one may deduce quantitative quantum dynamical lower bounds (using a general theory initiated by Guarneri \cite{g}, Combes \cite{c}, and Last \cite{l}).

We will be concerned with quantum dynamical upper bounds in the spirit of Killip, Kiselev, and Last \cite{kkl}. While there is no general method known to this date which bounds the ``fast'' part of the wavepacket time evolution from above, the authors of \cite{kkl} propose a method to at least bound the spreading of the ``slow'' part from above. They apply their method to the operator of the form \eqref{oper}, where $\lambda > 8$, $\alpha = (\sqrt{5} - 1)/2$, and $\theta = 0$. In this paper we consider the case of general $\alpha$ and $\theta$.

Before stating our main theorem, we fix some notation. Let $\delta_n$ be the element of $\ell^2(\Z)$ which is supported at $n \in \Z$ and obeys $\delta_n(n) = 1$. Given a function $\psi : \Z \rightarrow \C$ and $L > 0$, we define

\begin{equation}\label{psilnorm}
\| \psi \|_L^2 = \sum_{n = - \lfloor L \rfloor}^{\lfloor L \rfloor} | \psi (n) |^2 + (L - \lfloor L \rfloor ) \left( | \psi ( -\lfloor L \rfloor -1 ) |^2 + | \psi ( \lfloor L \rfloor + 1 ) |^2 \right).
\end{equation}
For a function $A : [0, \infty ) \rightarrow \R$, we define
$$
\langle A(t) \rangle_T = \frac{2}{T} \int_0^\infty e^{-2t/T} A(t) \, dt.
$$
Finally, for $\alpha \in (0,1)$ irrational, we consider its continued fraction expansion

\begin{equation}\label{contfrac}
\alpha = \cfrac{1}{a_1+ \cfrac{1}{a_2+ \cfrac{1}{a_3 + \cdots}}}
\end{equation}
with uniquely determined $a_k \in \N$. Let us define the associated rational approximants $p_k/q_k$ of $\alpha$ by

\begin{alignat}{3}
\label{pkdef}
p_0 &= 0, &\quad    p_1 &= 1,   &\quad  p_k &= a_k p_{k-1} + p_{k-2},\\
\label{qkdef}
q_0 &= 1, &     q_1 &= a_1, &       q_k &= a_k q_{k-1} + q_{k-2}.
\end{alignat}
The rational numbers $p_k/q_k$ are known to be best approximants to $\alpha$ and, in particular, $p_k$ and $q_k$ are relatively prime. See Khinchin \cite{khin} for background on continued fraction expansions.

We let

\begin{equation}\label{edef}
\mathcal{E} = \{ \alpha \in (0,1) \mbox{ irrational} : \exists  B \mbox{ such that } q_k + 1 \le B^k \, \forall k \in \N \}.
\end{equation}
It is known that $\mathcal{E}$ has full Lebesgue measure \cite{khin}.

Our main result reads as follows:

\begin{theorem}\label{main}
There are constants $C_1, C_2 > 0$ such that for every $\lambda > 20$, every $\alpha \in \mathcal{E}$, and every $\theta \in [0,1)$, we have

\begin{equation}\label{dynbound}
\left\langle \| e^{-itH_{\lambda, \alpha, \theta}} \delta_1 \|^2_{C_1 T^{p(\lambda,\alpha)}} \right\rangle_T \ge C_2
\end{equation}
for $T$ large enough {\rm (}i.e., for $T \ge T_0(\alpha,\theta)${\rm )}, where
$$
p(\lambda,\alpha) = \frac{6 \log B}{\log \left( \frac{\lambda - 8}{3} \right) }
$$
and $B$ is associated to $\alpha$ via \eqref{edef}.
\end{theorem}

\noindent\textit{Remarks.} (a) The physical interpretation of \eqref{dynbound} is the following. At any time $T \ge T_0(\theta)$, the averaged probability to find the particle in a ball of radius $C_1 T^{p(\lambda,\alpha)}$ is uniformly bounded away from zero. This gives an upper bound on the \textit{slow} part of the wavepacket time evolution.\\[1mm]
(b) Since for fixed $\alpha \in \mathcal{E}$, $p(\lambda,\alpha) \rightarrow 0$ as $\lambda \rightarrow \infty$, we see that the slow part of the wavepacket moves arbitrarily slowly if we choose large enough coupling.\\[1mm]
(c) This theorem was shown by Killip et al.\ in \cite{kkl} for the particular case $\alpha = (\sqrt{5} - 1)/2$ and $\theta = 0$. They only require $\lambda > 8$. For this particular rotation number $\alpha$, our extension to arbitrary phase $\theta$ works for the same range of $\lambda$-values.\\[1mm]
(d) Our extension to arbitrary phase $\theta$ is based on a fine analysis of the local combinatorial structure of the sequences $v_{\alpha,\theta}$ which might be of independent interest.\\[1mm]
(e) The key ingredient in our extension to almost every rotation number $\alpha$ is a recent result of Liu and Wen \cite{lw} which generalizes a paper of Raymond \cite{r} (which in turn is at the heart of the proof in \cite{kkl}) to arbitrary rotation number.\\[1mm]
(f) As in \cite{kkl}, the theorem, while stated for initial vector $\delta_1$, can be recast for other initial vectors from $\ell^2(\Z)$.\\[1mm]
(g) Alternatively, the bound \eqref{dynbound} holds for every $T$ if one allows for $\theta$-dependent $C_1$ and adjusted $p(\lambda,\alpha) = p(\lambda,\alpha,\theta)$.

\medskip

The organization of the article is as follows. In the next section, we recall important properties of the operators $H_{\lambda,\alpha,\theta}$ and in particular describe the results of Liu and Wen. In Section~3 we establish all the necessary ingredients for an application of the general method of Killip, Kiselev, and Last. Finally, in Section~4, we put everything together and prove Theorem~\ref{main}.

\section{Spectra of Periodic Approximants and a Result of Liu and Wen}

In this section we discuss the structure of the spectra of the standard periodic approximants to a given $H_{\lambda,\alpha,\theta}$ and describe recent results of Liu and Wen.

Fix $\lambda > 0$ and some irrational $\alpha \in (0,1)$. Let $(a_k)_{k \in \N}$ be the continued fraction expansion coefficents associated with $\alpha$ via \eqref{contfrac}, and let $p_k/q_k$ be the rational approximants to $\alpha$. Following \cite{bist,lw,r,s1}, we define the matrices $M_k(E) = M_k(E; \lambda,\alpha)$, for $k \ge 1$, by

\begin{equation}\label{mkdef}
M_k(E) = M(E,q_k,\lambda,\alpha,0),
\end{equation}
where, for $n \in \Z$,

\begin{equation}\label{transfer}
M(E,n,\lambda,\alpha,\theta) = \left\{ \begin{array}{cl} T(E,n,\lambda,\alpha,\theta) \times \cdots \times T(E,1,\lambda,\alpha,\theta) & \mbox{ if } n \ge 1 \\ 
\left( \begin{array}{cr} 1 & 0 \\ 0 & 1 \end{array} \right) & \mbox{ if } n = 0\\
T(E,n+1,\lambda,\alpha,\theta)^{-1} \times \cdots \times T(E,0,\lambda,\alpha,\theta)^{-1} & \mbox{ if } n \le -1
\end{array} \right.
\end{equation}
and, for $m \in \Z$,

\begin{equation}\label{tmdef}
T(E,m,\lambda,\alpha,\theta) = \left( \begin{array}{cr} E - \lambda v_{\alpha,\theta}(m) & -1 \\ 1 & 0 \end{array} \right).
\end{equation}
Thus, $M_k(E)$ is the standard transfer matrix associated with the operator $H_{\lambda,\alpha,\theta = 0}$ and the interval $[1,q_k]$.

For $k = -1$ and $k = 0$, we define

\begin{equation}\label{m-1def}
M_{-1}(E) = \left( \begin{array}{cr} 1 & -\lambda \\ 0 & 1 \end{array} \right) \; \mbox{ and } \; M_0(E) = \left( \begin{array}{cr} E & -1 \\ 1 & 0 \end{array} \right).
\end{equation}

Furthermore, we let

\begin{equation}\label{tkpdef}
t_{(k,p)}(E) = \tr \left( M_{k-1}(E) M_k(E)^p \right)
\end{equation}
and

\begin{equation}\label{skpdef}
\sigma_{(k,p)} = \{ E \in \R : |t_{(k,p)}(E)| \le 2 \}.
\end{equation}
The set $\sigma_{(k,p)}$ is the spectrum of a periodic Schr\"odinger operator whose transfer matrix over one period is given by $M_{k-1}(E) M_k(E)^p$. Consequently, it consists of a finite number ($= p q_k + q_{k-1}$, to be precise) of closed intervals (``bands''). In particular, the set $\sigma_{(k+1,0)}$ is the spectrum of the periodic Schr\"odinger operator $H_{\lambda, p_k/q_k,0}$ (i.e., its potential results from $\lambda v_{\alpha,0}$ by replacing $\alpha$ by the rational approximant $p_k/q_k$); compare \cite{bist}.

We recall two key results \cite{bist,r}:

\begin{equation}\label{mkrec}
M_{k+1}(E) = M_{k-1}(E) M_k(E)^{a_{k+1}},
\end{equation}
which implies $t_{(k+2,0)} = t_{(k,a_{k+1})}$, and

\begin{equation}\label{invariant}
t_{(k+1,0)}^2 + t_{(k,p)}^2 + t_{(k,p+1)}^2 - t_{(k+1,0)} t_{(k,p)} t_{(k,p+1)} = 4 + \lambda^2.
\end{equation}
 
We first discuss how the spectrum of $H_{\lambda,\alpha,\theta}$ is approximated by the sets $\sigma_{(k,p)}$. It is easy to see (and was noted in \cite{bist}) that the set $\sigma(H_{\lambda,\alpha,\theta})$ is independent of $\theta$, that is, there exists a compact set $\Sigma_{\lambda,\alpha} \subset \R$ such that
$$
\sigma(H_{\lambda,\alpha,\theta}) = \Sigma_{\lambda,\alpha} \; \mbox{ for every } \theta \in [0,1).
$$
Let us define the following three types of bands:

\begin{center}
\begin{tabular}{rcl}
$(k,I)$-type band & : & a band of $\sigma_{(k,1)}$ which is contained in a band of $\sigma_{(k,0)}$,\\
$(k,II)$-type band & : & a band of $\sigma_{(k+1,0)}$ which is contained in a band of $\sigma_{(k,-1)}$,\\
$(k,III)$-type band & : & a band of $\sigma_{(k+1,0)}$ which is contained in a band of $\sigma_{(k,0)}$.
\end{tabular}
\end{center}
These bands are mutually disjoint and they are called \textit{spectral generating bands of order k} by Liu and Wen. Their union is denoted by $\mathcal{G}_k$. The following was shown in \cite{lw}:

\begin{lemma}
{\rm (a)} $\Sigma_{\lambda,\alpha} = \cap_k \mathcal{G}_k$.\\
{\rm (b)} For $k > 0$, every spectral generating band of order $k+1$ is contained in some spectral generating band of order $k$.
\end{lemma}

From this lemma, we see that for every $k$, every energy in the spectrum lies in some spectral generating band of order $k$. Thus, we can assign to it a one-sided infinite sequence over the alphabet $\mathcal{A} = \{ I, II, III \}$. 

Define the matrices $T_m$ by
$$
T_m = (t_m(i,j))_{i,j = 1,2,3} = \left( \begin{array}{ccc} 0 & 1 & 0 \\ a_m + 1 & 0 & a_m \\ a_m & 0 & a_m -1 \end{array} \right).
$$
An $(m-1,i)$-type band generates $t_m (i,j)$ bands of $(m,j)$ type. Moreover, every spectral generating band of order $k$ is associated with a unique word $i_0 i_1 \ldots i_k \in \mathcal{A}^{k+1}$, called its \textit{type index}. 

A central result of \cite{lw}, whose proof makes critical use of \eqref{mkrec} and \eqref{invariant}, establishes lower bounds for the derivative with respect to the energy of the trace function $t_{(k,p)}$ for energies from spectral generating bands. To formulate this result, we need the following sequence of matrices:

\begin{equation}\label{pkmatdef}
P_m = (p_m(i,j))_{i,j = 1,2,3} = \left( \begin{array}{ccc} 0 & t_\lambda^{-(a_m - 1)} & 0 \\ a_m/t_\lambda & 0 & a_m/t_\lambda \\ a_m/t_\lambda & 0 & a_m/t_\lambda \end{array} \right),
\end{equation}
where
$$
t_\lambda = \frac{3}{\lambda - 8}.
$$

Then, Liu and Wen prove the following result (essentially Proposition~5 of \cite{lw}; the explicit estimate appears in the proof of this proposition).

\begin{prop}\label{lwthm}
Let $\lambda > 20$. For every spectral generating band $B$ with type index $i_0 i_1 \ldots i_k$, we have for every $E \in B$,
$$
|T'(E)| \ge p_1(i_0,i_1) p_2(i_1,i_2) \cdots p_k(i_{k-1},i_k),
$$
where $T$ is the appropriate trace function {\rm (}i.e., $t_{(k,p)}$ if $B$ is a band of $\sigma_{(k,p)}${\rm )}.
\end{prop}

We now put the above result in a form suitable for our purpose. Denote, for $k \in \N$,

\begin{equation}\label{agtone}
A_k = \Pi_{i=1}^k a_i \ge 1
\end{equation}
and
$$
x_k(E) = t_{(k+1,0)}(E) \; \mbox{ and } \; \sigma_k = \sigma_{(k+1,0)}.
$$
Then we have (cf.\ \cite{bist})

\begin{equation}\label{spectrumrep}
\Sigma_{\lambda,\alpha} = \bigcap_{k \in \N} (\sigma_k \cup \sigma_{k+1})
\end{equation}
and the following result for $x'(E)$:

\begin{prop}\label{lwcor}
Let $\lambda > 20$. Then for every $k \in \N$ and every $E \in \sigma_k$, we have
$$
|x_k'(E)| \ge A_k \xi(\lambda)^{k-1},
$$
where
$$
\xi(\lambda) = \left( \frac{\lambda - 8}{3} \right)^{1/2}.
$$
\end{prop} 

\begin{proof}
This is an immediate consequence of the results above. We note that every band of $\sigma_k$ is either a $(k,II)$-type band or a $(k,III)$-type band and hence generating. Moreover, we see from the specific form of the matrices $P_m$ in \eqref{pkmatdef} that we pick up a factor $(\lambda - 8)/3$ in at least every other step. Finally, for $\lambda > 20$, we have
$$
t_{\lambda}^{-(a_m - 1)} \ge a_m
$$
and we therefore pick up a factor $a_m$ in every step.
\end{proof}

\section{Bounds on Transfer Matrices and Variation of the Phase}

In this section we establish the input to KKL theory, which will then be used to prove Theorem~\ref{main} in the next section. Namely, we will study derivatives of traces of transfer matrices with respect to the energy, and we will then prove lower bounds for averaged transfer matrix norms. Our approach is combinatorial in nature. Namely, we will use the partition approach to Sturmian sequences, as developed in \cite{dl1}, to study the local structure of the sequences $v_{\alpha,\theta}$ for general $\alpha,\theta$. In particular, we will exhibit plenty of occurrences of words conjugate to words whose associated trace derivatives we can control, thanks to Propositions~\ref{lwthm} and \ref{lwcor}, which is crucial since the trace is invariant with respect to cyclic permutation. This will then allow us to use the results of Section~2 to study the issues at hand.

Let $M(E,n,\lambda,\alpha,\theta)$ be the transfer matrix as defined in \eqref{transfer}. We denote, for $L \ge 1$,
$$
\|M(E,\lambda,\alpha,\theta)\|_L^2 = \sum_{n=1}^{\lfloor L \rfloor - 1} \|M(E,n,\lambda,\alpha,\theta)\|^2 + ( L - \lfloor L \rfloor ) \|M(E, \lfloor L \rfloor,\lambda,\alpha,\theta)\|^2,
$$
where $\|M(E,n,\lambda,\alpha,\theta)\|$ is the usual operator norm of the matrix $M(E,n,\lambda,\alpha,\theta)$. While this notation conflicts somewhat with the definition in \eqref{psilnorm}, it is standard (and was used in \cite{kkl} and previous works). We hope that this does not lead to any real confusion.

As shown in \cite{kkl} (using a formula from \cite{toda}), we have the following estimate of this quantity in terms of the derivative of the trace:

\begin{equation}\label{keyestimate}
\frac{\partial}{\partial E} ( \tr M(E,L,\lambda,\alpha,\theta) ) \le 4 \|M(E,\lambda,\alpha,\theta)\|^3_{L+1}.
\end{equation}

Thus, combining this estimate with the result given in Proposition~\ref{lwthm} (resp., Proposition~\ref{lwcor}), we obtain lower bounds on $\|M(E,\lambda,\alpha,\theta)\|_L$ for energies in spectral generating bands (for suitable $L$ and, at this point, $\theta = 0$). 

The quantity $\|M(E,\lambda,\alpha,\theta)\|_L$ for $L \le 0$ is defined analogously. Clearly, an estimate similar to \eqref{keyestimate} holds.

Our immediate goal is to study the trace of $M(E,n,\lambda,\alpha,\theta)$ for general $\theta$. To this end, we analyze the local structure of the sequences $v_{\alpha,\theta}$. We show in particular that the traces of the transfer matrices over intervals of length $q_k$ exhibit a rather strong invariance property with respect to a variation of the phase.

Let us first recall some combinatorial notions. As general references, we want to mention \cite{loth1,loth2}. Let $A$ be a finite set, called the alphabet, and denote by $A^*,A^\N,A^\Z$ the set of finite, one-sided infinite, and two-sided infinite words over $A$, respectively. A word $v$ is called a subword (or factor) of some word $u$ if there are words $w_1,w_2$ (possibly empty) such that $u = w_1 v w_2$. If $w = w_1 \ldots w_m$ with $w_i \in A$, then the word $w^R = w_m \ldots w_1$ is called the reversal of $w$. A word $w$ with $w = w^R$ is called a palindrome. If a word $w$ can be written as $w = uv$ with words $u,v$, we call $u$ a prefix of $w$ and $v$ a suffix of $w$. Two words $w_1,w_2$ are said to be conjugate if there are words $u,v$ such that $w_1 = uv$ and $w_2 = vu$. This is equivalent to $w_2$ being equal to a cyclic permutation of $w_1$. Given any word $w$, we denote by $P_w$ the set of its finite subwords, and by $P_w(n)$ the set of its finite subwords of length $n$, $n \in \N$. Write $p_w(n)$ for the cardinality of $P_w(n)$; the function $p_w : \N \rightarrow \N$ is called the complexity function associated with $w$. If the word $w$ is infinite and uniformly recurrent (i.e., each of its finite subwords occurs infinitely often and with bounded gaps), we define the hull associated with $w$ by $\Omega_w = \{ s \in A^\Z : P_s = P_w \}$. 

In our concrete setting, the alphabet will be given by $A = \{0,1\}$ and the word $w$ will be given by the restriction of $v_{\alpha,0}$ to $\N$, that is,
$$
w = v_{\alpha,0} (1) v_{\alpha,0} (2) v_{\alpha,0} (3) \ldots \in \{0,1\}^\N.
$$
The word $w$ is called a characteristic Sturmian sequence and its combinatorial properties have been studied extensively. For example, it is well known that its complexity function is given by

\begin{equation}\label{complex}
p_w (n) = n+1 \mbox{ for every } n.
\end{equation}
This implies that for every $n$, $P_w(n)$ contains exactly one word which has two extensions to the right to form a word in $P_w(n+1)$; every other word extends uniquely to the right. This special word, let us denote it by $r_n$, is called the right-special factor of $w$ of length $n$ and, for later use, we note the following (Proposition~2.1.23 of \cite{loth2}):

\begin{equation}\label{rightspecial}
r_n = v_{\alpha,0} (n) \ldots v_{\alpha,0} (1).
\end{equation}
That is, the set of right-special factors of $w$ coincides with the set of reversals of its prefixes. Also for later use, we note that it follows from Proposition~4.5 of \cite{b2} that

\begin{equation}\label{palindrome}
\mbox{For every } k \in \N, \; v_{\alpha,0} (1) \ldots v_{\alpha,0} (q_k - 2) \mbox{ is a palindrome.}
\end{equation}

Moreover, it is also well known that $w$ is uniformly recurrent and that for every $\theta$, we have $v_{\alpha,\theta} \in \Omega_w$. In particular, we have 

\begin{equation}\label{samewords}
P_{v_{\alpha,\theta}} = P_w \mbox{ for every } \theta.
\end{equation}

Our goal is to study the sets $P_w(q_k)$, $k \in \N$. One element of $P_w(q_k)$ is certainly given by the prefix $s_k$ of $w$ of length $q_k$. As is well known, the words $s_k$, $k \in \N$ obey recursive relations. Since this fact is crucial to our proof, we recall this result briefly: 

\begin{equation}\label{skrec}
s_0 = 0, \; s_1 = 0^{a_1 - 1} 1, \; s_k = s_{k-1}^{a_k} s_{k-2} \mbox{ for } k \ge 2.
\end{equation}
Note that this is where \eqref{mkrec} comes from.

Our first observation is very simple:

\begin{lemma}\label{lastsymb}
For $k \ge 2$, the word $s_k$ has suffix $10$ if $k$ is even and it has suffix $01$ if $k$ is odd.
\end{lemma}

\begin{proof}
This follows immediately from the recursion \eqref{skrec}.
\end{proof}

Define $\overline{\cdot} : A \rightarrow A$ by $\overline{0} = 1$, $\overline{1} = 0$. Our next goal is to list all elements of $P_w(q_k)$ explicitly. Write $s_k = s_k^{(1)} \ldots s_k^{(q_k)}$ with $s_k^{(i)} \in A$, $1 \le i \le q_k$.

\begin{lemma}\label{list}
For every $k \ge 0$, the $q_k + 1$ elements of $P_w(F_k)$ are given by

\begin{itemize}
\item The $q_k$ cyclic permutations of $s_k$ which are mutually distinct, and
\item the word $\overline{s_k^{(q_k)}} s_k^{(1)} \ldots s_k^{(q_k - 1)}$.
\end{itemize}
\end{lemma}

\begin{proof}
As a preliminary remark, we note that it follows from \eqref{skrec} that $w$ contains the factor $s_{k-1} s_k s_k$, for every $k \ge 1$. Namely, for $k \ge 1$, $w$ has the prefix

\begin{eqnarray*}
s_{k+3} & = & s_{k+2}^{a_{k+3}} s_{k+1}\\
& = & \left( s_{k+1}^{a_{k+2}} s_k \right)^{a_{k+3}} s_k^{a_{k+1}} s_{k-1}\\
& = & \left( \left( s_k^{a_{k+1}} s_{k-1} \right)^{a_{k+2}} s_k \right)^{a_{k+3}} s_k^{a_{k+1}} s_{k-1},
\end{eqnarray*}
which contains the factor $s_{k-1} s_k s_k$.

First of all, the claim of the proposition can be verified easily for $k = 0$ and $k = 1$. We therefore consider the case $k \ge 2$ and write $a$ for the rightmost symbol of $s_k$. Then, using Lemma~\ref{lastsymb}, we have the following structure somewhere in $w$:
$$
\begin{array}{ccc}
s_{k-1} & s_k & s_k \\
\fbox{\hspace{1cm} $| \, \overline{a}$} & \fbox{\hspace{1.6cm} $| \, a$} &\fbox{\hspace{1.6cm} $| \, a$} 
\end{array}
$$
In particular, all the words listed in the assertion of the lemma belong to $P_w(q_k)$. Finally, to conclude the proof all we have to show is that the $q_k$ cyclic permutations of $s_k$ are mutually distinct because by \eqref{complex} there are only $q_k + 1$ words in $P_w(q_k)$ and the list contains $q_k + 1$ words which are mutually distinct. Define, for $k \ge 0$, the height $h(s_k)$ of $s_k$ by $h(s_k) = $ number of $1$'s in $s_k$. It follows from the definition, \eqref{pkdef}, and \eqref{skrec} that $h(s_k) = p_k$. Since $p_k$ and $q_k$ are relatively prime for every $k \ge 0$, we get that for $k \ge 0$, $h(s_k)$ and $|s_k|$ are relatively prime. This implies that the cyclic permutations of $s_k$ are mutually distinct, for otherwise $s_k$ could be written as a power of some shorter word, contradicting the above observation.
\end{proof}

We see that there is only one word $b_k$ in $P_w(q_k)$ which is not a cyclic permutation of $s_k$ and it can be described explicitly. In particular, it follows from Lemma~\ref{lastsymb} and Lemma~\ref{list} that the following holds:

\begin{equation}\label{firstsym}
\mbox{The leftmost symbol of $b_k$ is } \left\{ \begin{array}{ll} 1 & \mbox{if $k$ is even,} \\ 0 & \mbox{if $k$ is odd,} \end{array} \right. 
\end{equation}
and 

\begin{equation}\label{lastsym}
\mbox{The rightmost symbol of $b_k$ is } \left\{ \begin{array}{ll} 1 & \mbox{if $k$ is even,} \\ 0 & \mbox{if $k$ is odd.} \end{array} \right. 
\end{equation}

Denote by $s_k^\theta$ the word $v_{\alpha,\theta} (1) v_{\alpha,\theta} (2) \ldots v_{\alpha,\theta} (q_k)$ and by $t_k^\theta$ the word $v_{\alpha,\theta} (-q_k + 1) v_{\alpha,\theta} (-q_k + 2) \ldots v_{\alpha,\theta} (0)$. We can now state the following combinatorial result:

\begin{prop}\label{cyclic}
For every $\theta$, we have that $s_k^\theta$ is a cyclic permutation of $s_k$ for all $k$ odd or for all $k$ even. The same statement is true for $t_k^\theta$.
\end{prop}

\begin{proof}
Fix $\theta$. If $v_{\alpha,\theta} (1) = 0$, then by \eqref{firstsym}, $s_k^\theta$ is conjugate to $s_k$ for all $k$ odd. Similarly, if $v_{\alpha,\theta} (1) = 1$, then by \eqref{firstsym}, $s_k^\theta$ is conjugate to $s_k$ for all $k$ even. A completely analogous argument, using \eqref{lastsym}, yields the claim for $t_k^\theta$.
\end{proof} 

This has the following immediate consequence for the traces of the transfer matrices. Let
$$
x_k(E,\lambda,\alpha,\theta) = \tr M(E,q_k,\lambda,\alpha,\theta).
$$

\begin{prop}\label{traceequalities}
For every $\lambda,\theta$, we have that 

\begin{equation}\label{traceeq}
x_k(E,\lambda,\alpha,\theta) = x_k(E,\lambda,\alpha,0) \mbox{ for every } E \in \R
\end{equation}
holds for all $k$ odd or for all $k$ even. In particular,

\begin{equation}\label{deriveq}
\frac{\partial}{\partial E} x_k(E,\lambda,\alpha,\theta) = \frac{\partial}{\partial E} x_k(E,\lambda,\alpha,0) \mbox{ for every } E \in \R
\end{equation}
holds for all $k$ odd or for all $k$ even. 
\end{prop}

\begin{proof}
This follows from Proposition~\ref{cyclic} and the invariance of the trace of a product with respect to a cyclic permutation of the factors.
\end{proof}

Similarly, we have the following result on the left half-line. Let
$$
y_k (E,\lambda,\alpha,\theta) = \tr M(E,-q_k,\lambda,\alpha,\theta).
$$

\begin{prop}\label{traceequalities2}
For every $\lambda,\alpha,\theta$, we have that 

\begin{equation}\label{traceeq2}
y_k(E,\lambda,\alpha,\theta) = x_k(E,\lambda,\alpha,0) \mbox{ for every } E \in \R
\end{equation}
holds for all $k$ odd or for all $k$ even. In particular,

\begin{equation}\label{deriveq2}
\frac{\partial}{\partial E} y_k(E,\lambda,\alpha,\theta) = \frac{\partial}{\partial E} x_k(E,\lambda,\alpha,0) \mbox{ for every } E \in \R
\end{equation}
holds for all $k$ odd or for all $k$ even. 
\end{prop}

\begin{proof}
We have

\begin{eqnarray*}
y_k (E,\lambda,\alpha,\theta) & = & \tr M(-q_k,E,\lambda,\alpha,\theta)\\
& = & \tr \left( T(-q_k + 1,E,\lambda,\alpha,\theta)^{-1} \times \cdots \times T(0,E,\lambda,\alpha,\theta)^{-1} \right)\\
& = & \tr \left( T(0,E,\lambda,\alpha,\theta) \times \cdots \times T(-q_k+1,E,\lambda,\alpha, \theta) \right),
\end{eqnarray*}
where in the last step we have used that the determinant is one and hence the trace is invariant with respect to inverting the matrix. By Proposition~\ref{cyclic} and invariance of the trace with respect to cyclic permutations, for all even $k$ or for all odd $k$, the right-hand side is equal to $x_k(E,\lambda,\alpha,0)$ for all $E$.
\end{proof}

For some fixed $\theta$, these results determine the traces $x_k(\theta)$ and $y_k(\theta)$ in terms of $x_k$ for one-half of the possible values of $k$, which leaves us with the question of how to investigate the other half. While it is certainly not true that for every $\theta$ and every $k$, $x_k(\theta)$ and $x_k$ are equal as functions of $E$, we will show that this is true at least for $k$ large enough and all but one $\theta$.

To this end, we will employ another combinatorial consideration which is based on the partitions of the sequences $v_{\alpha,\theta}$ introduced by D.~Lenz and the author in \cite{dl1}. 

\begin{definition}  
\rm  
Let $k \in \N_0$ be given. A $(k,\alpha)$-partition of a function $f:\Z \rightarrow \{0,1\}$ is a sequence of pairs $(I_j, z_j)$, $j\in\Z$ such that:  

\begin{itemize}  
\item the sets $I_j \subset \Z$ partition $\Z$;  
\item $1 \in I_0$;   
\item each block $z_j$ belongs to $\{s_k,s_{k-1}\}$; and  
\item the restriction of $f$ to $I_j$ is $z_j$. That is, $f_{d_j} f_{d_j +1} \ldots f_{d_{j+1}-1} = z_j$.  
\end{itemize}
Notice that $d_j$ is defined implicitly to be the left-hand endpoint of the interval $I_j$.
\end{definition}

We will suppress the dependence on $\alpha$ if it is understood to which $\alpha$ we refer. In particular, we will write $k$-partition instead of $(k,\alpha)$-partition. The sequences $v_{\alpha,\theta}$ have a unique decomposition property which is given in the following lemma (Lemma~3.2.(b) of \cite{dl1}).  
  
\begin{lemma}\label{partition-lemma}  
For every $k \in \N_0$, every irrational $\alpha \in (0,1)$, and every $\theta \in [0,1)$, there exists a unique $k$-partition $(I_j,z_j)$ of $v_{\alpha,\theta}$.  
\end{lemma}  

Using this partition lemma, we can now continue our study of the local structure of a given $v_{\alpha,\theta}$. Recall that $b_k$ denotes the unique word in $P_w(q_k)$ which is not conjugate to $s_k$. As a preliminary result, we show in the following lemma that $b_k$ can occur in $v_{\alpha,\theta}$ only at canonical positions relative to the $k$-partition of $v_{\alpha,\theta}$. Namely, whenever the $k$-partition yields the occurrence of some $s_{k-1}$ block followed by some $s_k$ block, we saw above that we get an occurrence of $b_k$. The following lemma says that these are the \textit{only} occurrences of $b_k$ in $v_{\alpha,\theta}$.

\begin{lemma}\label{synchron}
Let $\alpha \in (0,1)$ be irrational and let $\theta \in [0,1)$. If
$$
v_{\alpha,\theta}(m) \ldots v_{\alpha,\theta}(m + q_k - 1) = b_k
$$
for some $m \in \Z$, then $[m+1,m+q_k]$ is an interval $I_j$ belonging to the $k$-partition of $v_{\alpha,\theta}$. In particular, for every occurrence of $b_k$ in $v_{\alpha,\theta}$, we have the following local structure:

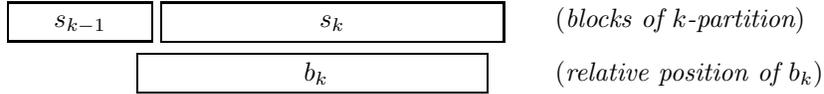
\begin{eqnarray*}
& \fbox{\rule{0cm}{2.1mm}\hspace{5mm}$s_{k-1}$\hspace{5mm}}  \; \fbox{\rule[-0.9mm]{0cm}{3mm}\hspace{2cm}$s_{k}$\hspace{2cm}} & \; \; \mbox{ {\rm (}blocks of $k$-partition{\rm )}}\\
& \hspace{15mm} \fbox{\hspace{21mm}$b_k$\hspace{20mm}} & \; \; \mbox{ {\rm (}relative position of $b_k${\rm )}}
\end{eqnarray*}
\end{lemma}

\begin{proof}
Recall that it follows from \eqref{complex} that there is exactly one factor of length $q_k - 1$ which does not have a unique extension to the right to a factor of length $q_k$, and it follows from \eqref{rightspecial} and \eqref{palindrome} that this factor is different from $v_{\alpha,\theta}(m+1) \ldots v_{\alpha,\theta}(m + q_k - 1)$. Thus this factor extends uniquely to the right, and we necessarily have $v_{\alpha,\theta}(m) \ldots v_{\alpha,\theta}(m + q_k) = s_k$. Now we can apply Lemma~3.3 of \cite{dl2} which says that this occurrence of $s_k$ must correspond to one from the $k$-partition. That this $s_k$ block must be preceded by an $s_{k-1}$ block in the $k$-partition is forced by $v_{\alpha,\theta} (m)$ being equal to the first letter of $b_k$. 
\end{proof}

\begin{prop}
Let $\alpha \in (0,1)$ be irrational and let $\theta \in [0,1)$. If $\theta \not= 1 - \alpha$, we have

\begin{equation}\label{traceeqglob}
x_k(E,\lambda,\alpha,\theta) = x_k(E,\lambda,\alpha,0) \mbox{ for every } E \in \R, \, k \ge k_0(\theta)
\end{equation}
and

\begin{equation}\label{traceeqglob2}
y_k(E,\lambda,\alpha,\theta) = x_k(E,\lambda,\alpha,0) \mbox{ for every } E \in \R, \, k \ge k_1(\theta).
\end{equation}
As before, this gives \eqref{deriveq} and \eqref{deriveq2} for the respective $k$-ranges.
\end{prop}
 
\begin{proof}
Assume first that \eqref{traceeqglob} fails. Our goal is to show 

\begin{equation}\label{badphase}
\theta = 1 - \alpha.
\end{equation}
Fix some $k$ and consider the $k$-partition of $v_{\alpha,\theta}$. If $s_k^{\theta}$ is conjugate to $s_k$, we have $x_k(E,\lambda,\alpha,\theta) = x_k(E,\lambda,\alpha,0)$ for every $E \in \R$. Conversely, if $s_k^{\theta}$ is not conjugate to $s_k$, then it follows from Lemma~\ref{synchron} that we must have the following situation:

\begin{equation}\label{badpic}
\begin{array}{rl}
\fbox{\rule{0cm}{2.1mm}\hspace{5mm}$s_{k-1}$\hspace{5mm}} & \fbox{\rule[-0.9mm]{0cm}{3mm}\hspace{2cm}$s_k$\hspace{2cm}} \\
\ldots 1 & 2 \ldots
\end{array}
\end{equation}
That is, the site $1$ is the right endpoint of $I_0$. If \eqref{traceeqglob} fails, then we have the situation depicted in \eqref{badpic} for infinitely many values of $k$. In other words, $v_{\alpha,\theta}$ restricted to $[2,\infty)$ coincides with $v_{\alpha,0}$ restricted to $[1,\infty)$ because as a one-sided infinite word, it has infinitely many $s_k$'s as prefixes. Since the phase $\theta$ can be recovered uniquely from $v_{\alpha,\theta}$ restricted to a half-line (and $v_{\alpha,1 - \alpha}$ restricted to $[2,\infty)$ coincides with $v_{\alpha,0}$ restricted to $[1,\infty)$), we obtain \eqref{badphase}.\\[1mm]
Assume now that \eqref{traceeqglob2} fails. By Lemma~\ref{synchron}, we have the following situation for infinitely many values of $k$:

\begin{equation}\label{badpic2}
\begin{array}{rrl}
\fbox{\rule{0cm}{2.1mm}\hspace{4mm}$s_{k-1}$\hspace{4mm}} & \fbox{\rule[-0.9mm]{0cm}{3mm}\hspace{17mm}$s_k$\hspace{17mm}} & \fbox{\rule{0cm}{2.1mm}\hspace{12mm}$s_{k-1}$ or $s_k$\hspace{12mm}} \\
\ldots -q_k + 1 & \ldots 0 1 & 2 \ldots
\end{array}
\end{equation}
Again, the site $1$ is the right endpoint of $I_0$ and we conclude as above that \eqref{badphase} holds. 
\end{proof}

Next, we provide power-law lower bounds for $\|M(E,\lambda,\alpha,\theta)\|_L$ for $\lambda > 20$ and $\alpha \in \mathcal{E}$. This will be achieved using \eqref{keyestimate} and the results from the preceding section.

\begin{prop}
Let $\lambda > 20$ and $\alpha \in \mathcal{E}$. Then there exist constants $C , \zeta > 0$ such that for every $\theta \in [0,1)$ and every $E \in \Sigma_{\lambda,\alpha}$, we have

\begin{equation}\label{tmlowerb}
\|M(E,\lambda,\alpha,\theta)\|_L \ge C |L|^\zeta \; \mbox{ for } |L| \ge L_0(\theta).
\end{equation}
\end{prop}

\begin{proof}
For each $\alpha \in \mathcal{E}$, there is a constant $B$ such that the associated sequence $(q_k)_{k \in \N}$ obeys

\begin{equation}\label{expobound}
q_k + 1 \le B^k \; \mbox{ for every } k \in \N.
\end{equation}
Now let $\lambda > 20$, $\theta \in [0,1) \setminus \{ 1 - \alpha \}$, and $k \in \N$ such that

\begin{equation}\label{klarge}
k \ge \max \{ k_0(\theta), k_1(\theta) \},
\end{equation}
with $k_0(\theta)$ and $k_1(\theta)$ from \eqref{traceeqglob} and \eqref{traceeqglob2}. Let $E \in \Sigma_{\lambda,\alpha}$. Then, by \eqref{spectrumrep}, $E \in \sigma_k \cup \sigma_{k+1}$. Hence, by Proposition~\ref{lwcor}, we have either
$$
\frac{\partial}{\partial E} x_k(E,\lambda,\alpha,0) \ge A_k \xi(\lambda)^{k-1}
$$
or
$$
\frac{\partial}{\partial E} x_{k+1}(E,\lambda,\alpha,0) \ge A_{k+1} \xi(\lambda)^{k}.
$$
In either case, by \eqref{agtone}, \eqref{keyestimate}, and \eqref{traceeqglob} (yielding \eqref{deriveq}), we obtain
$$
4 \|M(E,\lambda,\alpha,\theta)\|_{q_{k+1}+1}^3 \ge \xi(\lambda)^{k-1}.
$$
Thus, if we consider $q_{k+1} + 1 \le L \le q_{k+2} + 1$ for $k$ obeying \eqref{klarge}, we get

\begin{eqnarray*}
\|M(E,\lambda,\alpha,\theta)\|_L & \ge & \|M(E,\lambda,\alpha,\theta)\|_{q_{k+1} + 1}\\
& \ge & \left( \frac{1}{4} \xi(\lambda)^{k-1} \right)^{1/3}\\
& \ge & C B^{\zeta (k+2)} \\
& \ge & C (q_{k+2} + 1)^\zeta\\
& \ge & C L^\zeta,
\end{eqnarray*}
with $B$ from \eqref{expobound} and, essentially,
$$
C = \frac{1}{4^{1/3}} \; \mbox{ and } \; \zeta = \frac{\log \xi(\lambda)}{3 \log B}
$$
(more precisely, $\zeta = \frac{\tilde{k} - 1}{\tilde{k} + 2} \cdot \frac{\log \xi(\lambda)}{3 \log B}$ with $\tilde{k} = \max \{ k_0(\theta), k_1(\theta) \}$, and we obtain an exponent $\zeta$ which is arbitrarily close to $\frac{\log \xi(\lambda)}{3 \log B}$ if we increase $\tilde{k}$ suitably).

A completely analogous proof, using \eqref{traceeqglob2} (yielding \eqref{deriveq2}), proves the claimed bound for negative $L$.

We are left with the exceptional case $\theta = 1 - \alpha$. Since $v_{\alpha,1-\alpha}$ is obtained from the sequence $v_{\alpha,0}$ by a unit shift to the right, there is a constant $C$, depending only on $\lambda$, such that for every $n \ge 1$, every $E \in \Sigma_{\lambda,\alpha}$, we have

\begin{align*}
\|M(&E,n,\lambda,\alpha,1-\alpha)\| = \\
& = \| T(E,n,\lambda,\alpha,1 - \alpha) \times \cdots \times T(E,1,\lambda,\alpha,1 - \alpha) \| \\
& = \| T(E,n-1,\lambda,\alpha,0) \times \cdots \times T(E,0,\lambda,\alpha,0) \| \\
& = \left\| T(E,n,\lambda,\alpha,0)^{-1} ( T(E,n,\lambda,\alpha,0) \times \cdots \times T(E,1,\lambda,\alpha,1) ) T(E,0,\lambda,\alpha,0) \right\| \\
& \ge \| T(E,n,\lambda,\alpha,0) \|^{-1} \cdot \| M(E,n,\lambda,\alpha,0) \| \cdot \| T(E,1,\lambda,\alpha,0)^{-1} \|^{-1} \\
& \ge C \|M(E,n,\lambda,\alpha,0)\|,
\end{align*}
and a similar inequality for $n \le 0$. This permits us to deduce \eqref{tmlowerb} for $\theta = 1 - \alpha$ from \eqref{tmlowerb} for $\theta = 0$ with the constant adjusted accordingly.
\end{proof}

\section{Application of KKL Theory}

In this section we prove Theorem~\ref{main} by combining \eqref{tmlowerb} with the general theory of \cite{kkl}. To this end, we first recall a result from \cite{kkl} and we then show how it applies in our concrete situation.

Consider a discrete, one-dimensional Schr\"odinger operator
$$
Hu(n) = u(n+1) + u(n-1) + V(n) u(n)
$$
in $\ell^2(\Z)$, define the transfer matrices $M(n,E)$ as usually, and define for $\varepsilon > 0$, the characteristic length scales $\tilde{L}^{\pm}_\varepsilon (E)$ (where $\pm \tilde{L}^{\pm}_\varepsilon (E) > 0$) by
$$
\|M(E)\|_{\tilde{L}^{\pm}_\varepsilon (E)} = 2 \| M(1,E)^{-1}\| \varepsilon^{-1}.
$$
Denote the spectral measure of the pair $(H,\delta_1)$ by $\mu$. Then, the following was shown in \cite{kkl} (essentially, Theorem~1.5 of that paper):

\begin{theorem}[Killip, Kiselev, Last]\label{kkltheo}
For any $T > 0$, $L > 2$, we have
$$
\langle \| e^{-itH} \delta_1 \|_L^2 \rangle_T > C \mu ( \{ E : \tilde{L}^{\pm}_{T^{-1}}(E) \le L \}),
$$
where $C$ is a universal positive constant.
\end{theorem}

We are now in a position to give the 

\begin{proof}[Proof of Theorem~\ref{main}.]
Consider the case of $\tilde{L}^{+}_{T^{-1}} (E)$ (the other one is analogous). The length scale $\tilde{L}^{+}_{T^{-1}} (E)$ is determined by
$$
\|M(E,\lambda,\alpha,\theta)\|_{\tilde{L}^{+}_{T^{-1}} (E)} = 2 \| M(E,1,\lambda,\alpha,\theta)^{-1}\| T.
$$
Under the assumptions of the theorem on $\lambda,\alpha$, we get from \eqref{tmlowerb} that for $T$ large enough, and every $E \in \Sigma_{\lambda,\alpha}$,
$$
\tilde{L}^{+}_{T^{-1}} (E)^{\frac{\log \xi(\lambda)}{3 \log B}} \le \tilde{C} T
$$
and hence
$$
\tilde{L}^{+}_{T^{-1}} (E) \le C T^{\frac{3 \log B}{\log \xi(\lambda)}}.
$$
This, together with Theorem~\ref{kkltheo}, implies the statement of Theorem~\ref{main} and concludes the proof.
\end{proof}

Let us comment on some of the remarks that are listed after the statement of Theorem~\ref{main}. First of all, if the rotation number $\alpha$ is equal to $(\sqrt{5}-1)/2$, then an analog of Proposition~\ref{lwthm}, sufficient for our purpose, was shown in \cite{kkl,r} for $\lambda > 8$. Since this result is the only place in our proof where we have to put a restriction on $\lambda$, our extension to every phase for this particular value of $\alpha$ works for every $\lambda > 8$, rather than every $\lambda > 20$.

Secondly, if one wants to establish the bound \eqref{dynbound} for every $T$, one needs \eqref{tmlowerb} for every $L$ which can easily be achieved for every \textit{fixed} $\theta$ by adjusting the constants $C$ and $\zeta$ suitably.

\end{document}